\documentstyle[11pt,aaspp4,epsfig,psfig]{article}

\def\deg{^\circ} 
\def\ltsima{$\; \buildrel < \over \sim \;$}
\def\simlt{\lower.5ex\hbox{\ltsima}} 
\def\gtsima{$\; \buildrel > \over \sim \;$} 
\def\simgt{\lower.5ex\hbox{\gtsima}} 

\def\bfC{{\rm C}} 
\def\bfR{{\rm R}} 
\def\bfS{{\rm S}} 
  
\def\bfs{{\bf s}}

\def\etal{{\rm et al.~\/}} 
\def\kms{\ifmmode {\rm \ km \ s^{-1}}\else$\rm km s^{-1}$\fi} 
\def\mpc{$\ {h^{-1}\rm Mpc}$} 
\def\bfv{{\bf v}}

\def\si{{\sigma_{8,{\rm iras}}}} 
\def\s8{{\sigma_8}} 
\def\h{{h}}
\def\ns{{n_{\rm s}}} 
\def\alphacmb{{\vec{\alpha}}_{\rm cmb}}
\def\alphairas{{\vec{\alpha}}_{\rm iras}}
\def\alphajoint{{\vec{\alpha}}_{\rm joint}}
\def\omegam{{\Omega_{\rm m}}}
\def\omegacdm{{\Omega_{\rm cdm}}} 
\def\omegab{{\Omega_{\rm b}}}
 
\def\omegal{{\Omega_\Lambda}}

\def\ns{{n_{\rm s}}} 
\def\omegabh2{{\omegab h^2}} 
\def\bi{{b_{\rm iras}}}
\def\betai{{\beta_{\rm iras}}}

\lefthead{Webster et al.}  \righthead{Joint Estimation of Cosmological
Parameters}

\begin{document}
\title{Joint estimation of cosmological parameters from CMB and IRAS data}

\author{Matthew Webster\altaffilmark{1}} 
\affil{Institute of Astronomy, Madingley Road, Cambridge CB3 0HA, UK}
\author{S.L. Bridle, M.P. Hobson and A.N. Lasenby} 
\affil{Astrophysics Group, Cavendish Laboratory,  Madingley Road, 
Cambridge CB3 0HE, UK}
\author{Ofer Lahav} \affil{Institute of Astronomy, Madingley Road,
Cambridge CB3 0HA, UK}
\and
\author{Gra\c ca Rocha} \affil{Department of Physics, Kansas State
University,  Manhattan, KS 66506, USA}
\altaffiltext{1}{email: mwebster@ast.cam.ac.uk}


\begin{abstract}
Observations of large scale structure (LSS) and the Cosmic Microwave
Background (CMB) each place separate constraints on the values of
cosmological parameters. We calculate a joint likelihood based on
various CMB experiments and the IRAS 1.2Jy galaxy redshift survey and
use this to find an overall optimum with respect to 
the free parameters. Our formulation self-consistently
takes account of the underlying mass distribution, which affects both
the CMB potential fluctuations and the IRAS redshift distortion. This
not only allows more accurate parameter estimation, but also removes
the parameter degeneracy which handicaps calculations based on either
approach alone. The family of Cold Dark Matter (CDM)
models analysed corresponds to a
spatially-flat universe with an initially scale-invariant
spectrum and a cosmological constant. Free parameters in the joint
model are the mass density due to all matter ($\omegam$), 
Hubble's parameter ($h = H_0 / 100\kms {\rm Mpc}^{-1}$), 
the quadrupole normalisation of the CMB power spectrum ($Q$) in $\mu$K,
and the IRAS light-to-mass bias ($\bi$).
Throughout the analysis, the baryonic density ($\Omega_b$) is
required is to satisfy the nucleosynthesis constraint
$\Omega_b h^2 = 0.024$. Results from the two
data sets show good agreement, and the joint optimum lies at 
$\omegam = 0.39$, $h = 0.53$, $Q = 16.96$ $\mu$K, and $\bi = 1.21$.
The 68 per cent confidence intervals are:
$0.29<\omegam<0.53$, $0.39<h<0.58$, $15.34<Q<17.60$, and $0.98<\bi<1.56$.
For the above parameters the normalisation and shape of the mass power-spectrum 
are $\sigma_8=0.67$ 
and $\Gamma=0.15$, 
and the age of the Universe is $16.5$~Gyr.

\end{abstract}

\keywords{Cosmic microwave background, Large-scale structure}


%

\section{Introduction}

Astronomical observations allow us to evaluate cosmological models,
and determine likely values for the parameters within them. As the
variety and depth of these observations have improved, so too have the
techniques for comparing them against theoretical predictions.

Observations of anisotropies in the Cosmic Microwave Background (CMB)
provide one of the key constraints for cosmological models and a
significant quantity of experimental data already exists. By comparing
the power spectrum of the CMB fluctuations derived from these
experiments with the power spectra predicted by different cosmological
models it is possible to set constraints on the value of certain
cosmological parameters (e.g. \cite{sh97}, \cite{lineweaver97},
and references therein).

Galaxy redshift surveys, mapping large scale structure (LSS), provide
another cosmologically important set of observations.  The clustering
of galaxies in redshift-space is systematically different from that in
real-space (\cite{kaiser87}). The mapping between the two is a
function of the underlying mass distribution, in which the galaxies
are not only tracers, but also velocity test particles
(\cite{lahav96}).  Many techniques have been developed for estimating
this mapping (\cite{yahil91,kaiser91}). Statistical quantities can be
generated for a given cosmology and these used to constrain model
parameters through comparison with survey data
(\cite{fsl94,cole95,ht95,fn96,willick97}).

Estimates derived separately from each of these two data sets have
problems with parameter degeneracy. In the analysis of LSS data, there
is uncertainty as to how well the observed light distribution traces
the underlying mass distribution. The light-to-mass linear bias, $b$,
introduced to account for this uncertainty, affects the value of many
central cosmological parameters, and makes any identified optimum
degenerate (\cite{sw95}). Similarly, on the basis of CMB data alone,
there is considerable degeneracy (\cite{bond95c}) between $\h = H_0 /
100 \kms {\rm Mpc}^{-1}$  and the energy density $\omegal$ due to the
cosmological constant (\cite{cpt92}). This leads to poor estimation of
the baryon ($\omegab$) and total mass ($\omegam$) densities
(\cite{lineweaver97}).

In this letter, we combine results from a bandpower approach covering
a range of CMB experiments, with a likelihood analysis of the IRAS
1.2Jy survey, performed in spherical harmonics.  
We present a self-consistent formulation of CMB and LSS parameter
estimation. In particular, our method expresses the effects of the
underlying mass distribution on both the CMB potential fluctuations
and the IRAS redshift distortion. This breaks the degeneracy inherent
in an isolated analysis of either data set, and places tight
constraints on several cosmological parameters. Indeed, it is
unsurprising that the two data sets are complementary, given that they
sample our universe at extreme ends of its evolution.
For simplicity, we restrict our attention to inflationary,
Cold Dark Matter (CDM) models, assuming a flat universe with linear,
scale-independent biasing.
Other recent studies which combine CMB and LSS include 
\cite{gs98} and  \cite{eht98}. 

\section{CMB Parameter Estimation}


\subsection{Experimental Data}

Since the discovery of CMB fluctuations by the COBE satellite
(\cite{smoot,bennett}), several other experiments have also measured
CMB anisotropies over a wide range of angular scales. These
experiments include ground-based beam-switching experiments such as
Tenerife (\cite{me96,nature94}), Python (\cite{python}), South Pole
(\cite{spole}) and Saskatoon (\cite{sask}); balloon-borne instruments
such as ARGO (\cite{argo}), MAX (\cite{max}), and MSAM
(\cite{msam1,msam2}) and the ground-based interferometers CAT
(\cite{cat}) and OVRO (\cite{ovro}).

\begin{figure*}
\centerline{\epsfig{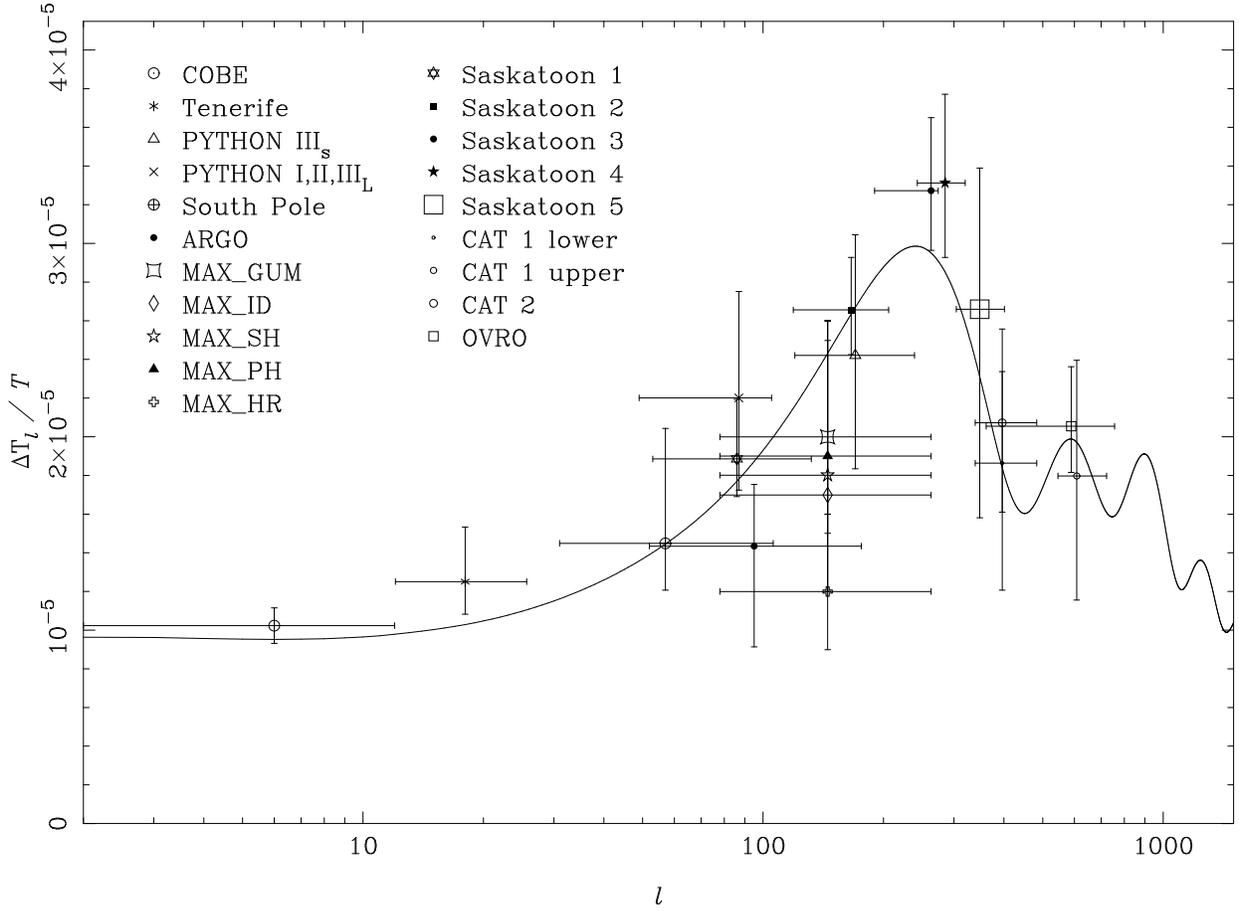}}
\caption{The data points used in the calculation of the CMB likelihood
function. The overlaid curve is a model evaluated with optimal parameters
shown in Table 1.}
\label{cmbdata}
\end{figure*}

These observations have resulted in a first estimate of the CMB power
spectrum and they are discussed in more detail by  \cite{sh97}.  In
particular, Hancock et al. display the window function $W_l$ for each
experiment and convert the level of anisotropy observed in each case
to flat bandpower estimates $(\Delta T_l/T) \pm \sigma$ centered on the
effective multipole  $l_{\rm eff}$ of the corresponding window
function (see below). The resulting CMB data points are plotted in
Fig.~\ref{cmbdata}, together with their 68 per cent confidence limits.
These confidence limits have been obtained using likelihood analyses
and hence incorporate uncertainties due to random errors, sampling
variance (\cite{samvariance}) and cosmic variance
(\cite{cosvariance2,cosvariance1}).  A discussion of possible
additional uncertainties due to contamination by foreground emission
is given by \cite{gr}.

The data points plotted in Fig.~\ref{cmbdata} differ slightly from
those given in \cite{sh97} as follows. The old Python point has been
replaced by two new points corresponding to the Py III$_S$ and
Py I, Py II and Py III$_L$ observations respectively (\cite{platt}).
The Tenerife point has been updated following the analysis of the full
two-dimensional data-set and a detailed treatment of atmospheric
effects (\cite{guti97}). The Saskatoon points have been
increased by 5 per cent, as suggested by the recent investigation of
systematic calibration errors in the Saskatoon experiment (Leitch,
private communication). We have added the results from the second CAT
field, reported in \cite{cat2} and also the OVRO point reported
by \cite{ovro}.


\subsection{Method}

Temperature fluctuations in the CMB are usually described in terms of
the spherical harmonic expansion
\begin{equation}
\frac{\Delta T(\theta,\phi)}{T}=\sum_{l=0}^\infty\sum_{m=-l}^l a_{l m}
\, Y_{l m}(\theta,\phi)\ ,
\end{equation}
from which we define the ensemble-average CMB power spectrum
$C_l=\langle |a_{l m}|^2 \rangle$. Alternatively, we may describe the
fluctuations in terms of their autocorrelation function
\begin{equation}
C(\theta)= \frac{1}{4\pi} \sum_{l=2}^\infty \, (2l + 1) \, C_l \,
P_l(\cos\theta)\ .
\end{equation}
The power in the CMB fluctuations observed by an experiment with
window function $W_l$ is then given by
\begin{equation}
C_{\rm obs}(0)= \left\langle{\left(\frac{\Delta T_{\rm
obs}}{T}\right)^2}\right\rangle = \frac{1}{4\pi} \sum_{l= 2}^\infty \,
(2l + 1) \, C_l \, W_l \ ,
\end{equation}
and for each experiment we define the flat bandpower by
\begin{equation}
{\frac{\Delta T_l}{T}} = \sqrt{{C_{\rm obs}(0)}\over{I(W_l)}} \ ,
\end{equation}
where $I(W_l)$ is defined (\cite{bond95a,bond95b}) as
\begin{equation}
I(W_l)=\sum_{l=2}^\infty \frac{(l+\textstyle{\frac{1}{2}})}
{l\,(l+1)}\ W_l \ .
\end{equation}
This flat bandpower estimate is centred on the effective multipole
\begin{equation}
l_{\rm eff}={{I(l \, W_l)}\over{I(W_l)}}\ .
\end{equation}

We wish to compare these bandpower estimates with those predicted by
different cosmological models. Varying the values of model parameters,
we calculate corresponding, predicted CMB power spectrum $C_l$, using
the Boltzmann code of \cite{seljak}.  This is
then used to calculate the predicted flat bandpower $\Delta T_l$ for
each experiment. The chi-squared statistic for a given set of
parameter values, $\alphacmb$, is then
\begin{equation}
\chi^2(\alphacmb)= \sum_{i=1}^{N_{\rm d}} {\frac{1}{\sigma_i^2}}
{{\left( \left[\Delta T_l^{\rm obs}\right]_i -\left[\Delta T_l^{\rm
pred}({\alphacmb})\right]_i \right)}^2} \ ,
\end{equation}
where $N_{\rm d}$ is the number of CMB data points plotted in
Fig.~\ref{cmbdata} (20 in this analysis).  Moreover, since the CMB
data points plotted in Fig.~\ref{cmbdata} were chosen such that no
two bandpower estimates come from experiments which observed
overlapping patches of sky and had overlapping window functions, we
may consider them as {\em independent} estimates of the CMB power
spectrum.
As the cosmic variance has already been  taken into account in deriving 
the flat bandpower estimates, the likelihood function 
is given simply by
${\cal L}_{\rm cmb} \propto e^{-\chi^2/2}$.

As mentioned above, we assume that the Universe is spatially flat, and
that there are no tensor contributions to the CMB power spectrum. We
take the primordial scalar perturbations to be described by the
Harrison-Zel'dovich power spectrum for which $\ns=1$, and further
assume that the optical depth to the last scattering surface is zero.

The normalisation of the CMB power spectrum is determined by $Q$,
which gives the strength of the quadrupole in $\mu K$, such that
\begin{equation}
C_2={{{4\pi}\over{5}}\,{{\left( {Q}\over{T_0} \right)}^2}} \ ,
\end{equation}
where $T_0$ is the average CMB temperature. The expansion rate of the
Universe is given by Hubble's parameter $h = H_0 / 100\kms{\rm Mpc}^{-1}$, while
$\omegacdm$ and $\omegab$ denote the density of the Universe in
CDM and the baryons respectively, each in units of the
critical density. Given that we assume a flat universe, but
investigate models where
\begin{equation}
\omegam \equiv (\omegacdm + \omegab) < 1 \ ,
\end{equation}
the shortfall is made up through a non-zero cosmological constant
$\Lambda$ such that
\begin{equation}
\omegal = 1 - \omegam =  {{\Lambda}\over{3{H_0^2}}} \ .
\end{equation}
Furthermore, we restrict our attention to models that satisfy the
nucleosynthesis constraint $\Omega_b h^2 = 0.024$ (\cite{tfb96}).
Thus we consider the reduced set of CMB parameters
\begin{equation}
\alphacmb \equiv \{Q, h, \omegacdm\} \ .
\end{equation}
In Section 4, we derive a set of joint parameters linking these to the
IRAS parameter set.


\section{IRAS Parameter Estimation}


\subsection{IRAS 1.2Jy Survey}

The IRAS surveys are uniform and complete down to Galactic latitudes
as low as $\pm 5\deg$ from the Galactic plane.  This makes them ideal
for estimating whole-sky density and velocity fields. Here, we use the
1.2 Jy IRAS survey (\cite{fisher95}),  consisting of 5313 galaxies,
covering 87.6\% of the sky with the incomplete regions being dominated by
the 8.7\% of the sky with $|b|<5\deg$.

In principle, the method we are using can be extended to account
explicitly for the incomplete sky coverage. However, we adopt the
simpler approach of smoothly interpolating the redshift distribution
over the missing areas (\cite{yahil91}).  The effects of this
interpolation on the computed harmonics have been shown to be
negligible (\cite{lahav94}).


\subsection{Method}

In this letter, we assume linear, scale-independent biasing, where
$\bi$ measures the ratio between fluctuations in the IRAS galaxy
distribution and the underlying mass density field:
\begin{equation}
{\left({{\delta\rho}/{\rho}}\right)}_{\rm iras} =
\bi\,{\left({{\delta\rho}/{\rho}}\right)_{\rm m}} \ .
\end{equation}
We note that biasing may be 
non-linear, stochastic, non-local, scale-dependent, epoch-dependent 
and type-dependent (e.g.  \cite{dl98}, \cite{tp98}, \cite{pen98},
\cite{bagla98}, \cite{blanton98}, and references therein).
For a linear  bias parameter, $\bi$,
the velocity and density fields in linear theory (\cite{peebles80})
are linked by a proportionality factor 
$\betai \equiv {{\omegam^{0.6}}/{\bi}}$, such that
\begin{equation}
\nabla\cdot\bfv =  -H_0\, \betai \,
\left({{\delta\rho}/{\rho}}\right)_{\rm m} 
\end{equation}

Statistically, the fluctuations in the real-space galaxy distribution
can be described by a power spectrum, ${P_\bfR(k)}$, which is
determined by the {\it{rms}} variance in the observed galaxy field,
measured for an 8\mpc\ radius sphere ($\si$) and a shape parameter
(e.g.  $\Gamma$ in equation~\ref{eqngamma}). The observed $\si$ is
related to the underlying $\s8$ for mass through the bias parameter,
such that $\si = \bi\,\s8$.

The approach we use in this letter follows  Fisher, Scharf \& Lahav
(1994; hereafter FSL),  and we include here only a brief introduction
to our technique. FSL provides a detailed description of the spherical
harmonic approach to parameter estimation.
A flux-limited, redshift-space density field can be decomposed into
spherical harmonics $Y_{lm}$, with coefficients
\begin{equation}
a_{lm}^\bfS = \sum\limits_{i=1}^{N_{\rm g}} f(s_i)\, Y_{lm} \left(
\hat\bfs_i \right) \ ,
\end{equation}
where $N_{\rm g}$ is the number of galaxies in the survey, and $f(s)$
is an arbitrary radial weighting function---this process is analogous
to Fourier decomposition, but instead using spherical basis
functions. The sum over galaxies in this equation can be rewritten as
a continuous integral of the density fluctuation field $\delta(\bfs)$
over redshift-space:
\begin{equation}
a_{lm}^\bfS = \int  d^3\bfs\,\,  \phi(r)\,  f(s)\,  \left[ 1 +
\delta_S(\bfs) \right] \, Y_{lm}(\hat\bfs) \ ,
\end{equation}
where $\phi(r)$ is the radial selection function of the survey,
evaluated at the real-space distance of the $i^{\rm th}$ galaxy.
 
As detailed in FSL, assuming the perturbations introduced by peculiar
velocities are small, the expected linear theory values for the
harmonic coefficients are
\begin{equation}
{\left\langle {\left| a_{lm}^\bfS \right|}^2 \right\rangle}_{\rm LT} =
{2\over\pi}\, { \int\limits_0^\infty  dk\,\, {k^2}\,  {P_\bfR(k)}\, {{
\left| {\Psi_l^\bfR(k) + \betai\Psi_l^\bfC(k)} \right|}^2} } \ .
\label{almeqn}
\end{equation}
Here, $\Psi^\bfR$ is the real-space window function, while
$\betai\Psi^\bfC$ is a ``correction'' term which embodies the redshift
distortion. 

For a given set of cosmological parameters, redshift-space harmonics
$a^j_{lm}$ can be calculated for different weighting functions
$f^j(r)$, ${j = 1,...,N}$. If the underlying density field is
Gaussian, on the basis of the coefficients predicted in 
equation~\ref{almeqn}, the likelihood of the survey harmonics can be
calculated as
\begin{equation}
{\cal L}_{\rm iras} \propto  {{\left| {\bf A} \right|}^{-{{1}\over{2}}}}\,
{{\rm exp} \left(
{{-{{1}\over{2}}}\left[{{\vec{a}}^{\rm T}}\,{\bf A}^{-1}\,{\vec{a}}\right]}
\right)} \ .
\label{iraslike}
\end{equation}
Here $\vec{a}$ is the vector of observed harmonics for different
shells and ${\bf A}$ is the corresponding covariance matrix, which
depends on the predicted harmonics. 
In addition to the ${\langle | a_{lm}^\bfS | \rangle}_{\rm
LT}$ from equation~\ref{almeqn}, 
these predicted harmonics will also have a shot-noise contribution
${\langle | a_{lm}^\bfS | \rangle}_{\rm SN}$ due to the discreteness
of survey galaxies.
Note that the argument of the exponent in equation~\ref{iraslike}
is simply 
$(-\chi^2/2)$, and that here the normalisation of the likelihood
function does depend on the free parameters (unlike in the CMB 
likelihood function). 

In this letter, we follow FSL and use four Gaussian windows centered
at 38, 58, 78, and 98 \mpc\ each with a dispersion of 8 \mpc. For each
window, we compute the corresponding weighted redshift
harmonics for the IRAS 1.2Jy catalog, and use these to determine the
likelihood of a given set of parameter values. Since our analysis is
valid only in the linear regime, we restrict the likelihood
computation to $l_{\rm max} = 10$ (corresponding to 120 degrees 
of freedom). Hence, the IRAS likelihood function
has a parameter vector 
\begin{equation}
\alphairas \equiv \{\betai, \si, \Gamma\} \ .
\end{equation}
Again, the linkage between these and the CMB parameters is discussed 
in section 4 below.


\section{Joint analysis}

Given the large number of parameters available between the two models,
it is important both to find links for joint optimisation, and to
decide which parameters can be frozen. From section 3, we have six
variables between the two models: \{$Q$, $\h$, $\omegacdm$,
$\betai$, $\si$, $\Gamma$\}. These can be reduced further by
expression in terms of core cosmological parameters. The IRAS
normalisation can be calculated as $\s8 \equiv f(\omegam, Q,
\Gamma)$ (\cite{bbks86,ebw92}), while the CDM shape parameter is
well approximated (\cite{sugiyama95}) by
\begin{equation}
\Gamma = \omegam\,\h \ {\rm exp}\left({-\omegab\left[1 +
{{\sqrt{\h/0.5}}\over{\omegam}}\right]}\right) \ .
\label{eqngamma}
\end{equation}
Meanwhile, we have shown above that 
$\omegam = \omegacdm + \omegab$, $\betai = \Omega_{\rm m}^{0.6}/\bi$, 
and $\si = \s8\bi$. Hence, the final, joint parameter space is
\begin{equation}
\alphajoint \equiv \{\h, Q, \omegam, \bi\}\ .
\label{alphajoint}
\end{equation}
As the IRAS and CMB probe very different scales and hence are 
assumed to be uncorrelated,  
the joint likelihood is given by
\begin{equation}
{\rm ln}\left({\cal L}_{\rm joint}\right) = 
{\rm ln}\left({\cal L}_{\rm cmb}\right) + 
{\rm ln}\left({\cal L}_{\rm iras}\right) \ .
\label{jointlike}
\end{equation}
%


\section{Results}
\label{results}

\begin{figure*}
\centering
{\epsfig{file=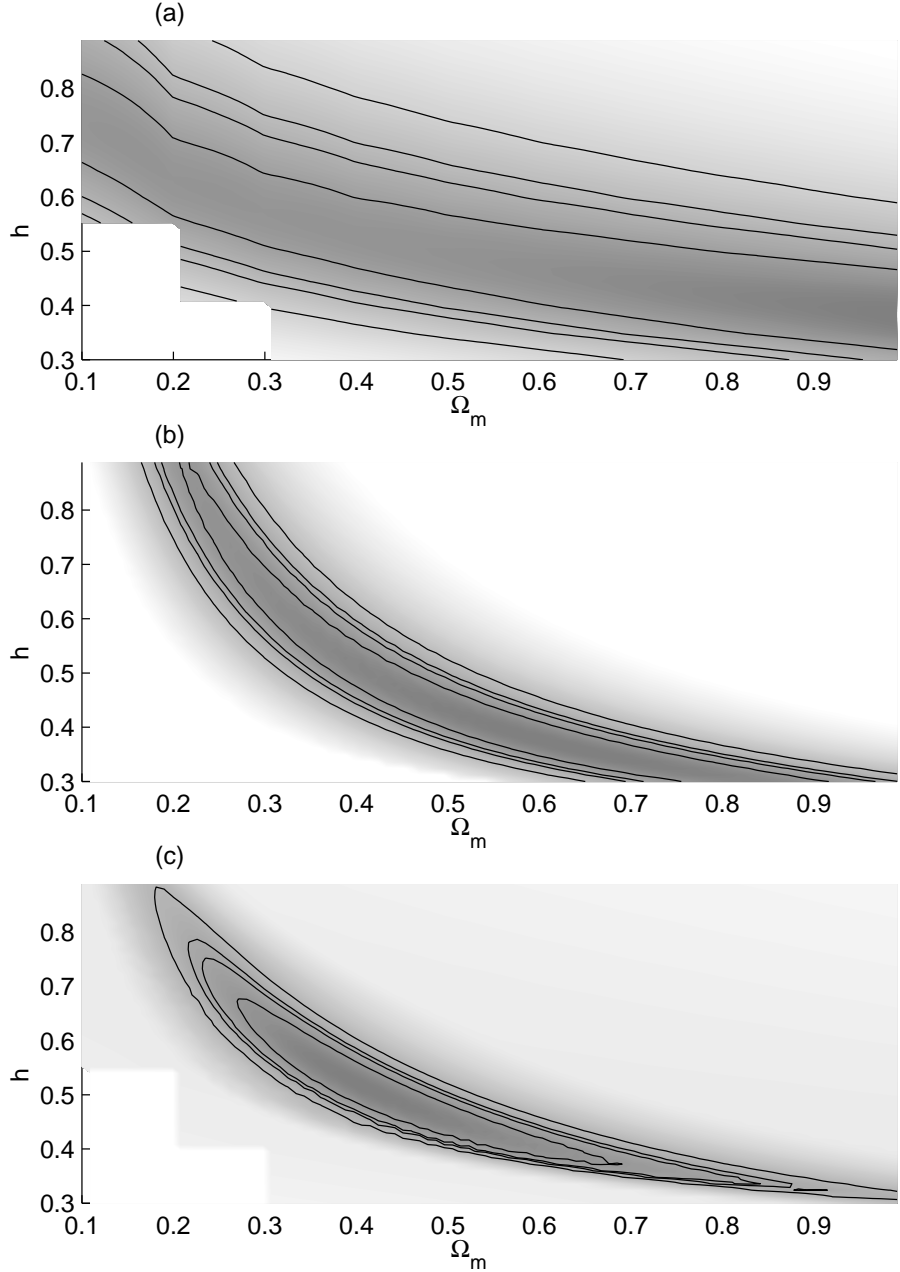,width=12cm}}
\caption
{The likelihood function in the \{$\omegam$, $\h$\}-plane, after
marginalisation over $Q$ and $\bi$, for (a) CMB alone, (b) IRAS alone
and (c) joint CMB and IRAS. 
The contours denote the 68, 90,
95 and 99 per cent confidence regions.}
\label{results1}
\end{figure*}

\begin{figure*}
\centering
{\epsfig{file=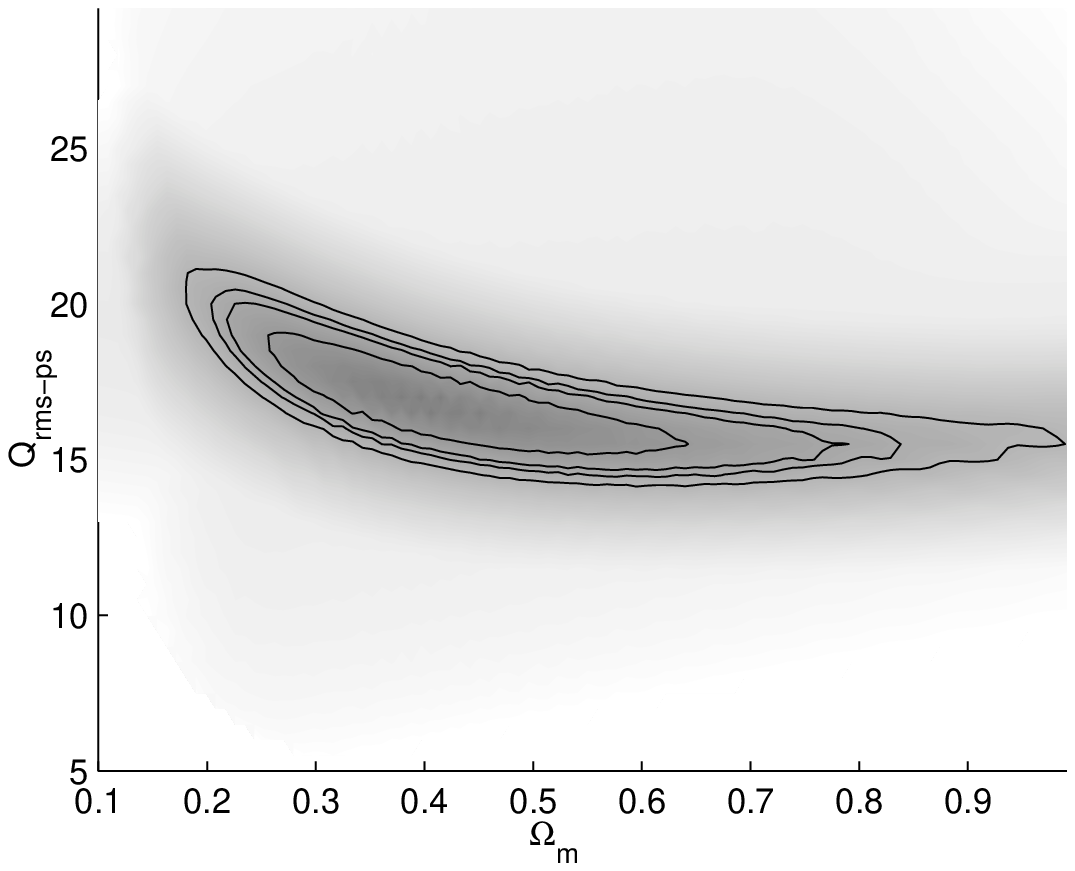,width=12cm}}
\caption
{The joint CMB+IRAS likelihood function in the
\{$\omegam$, $Q$\}-plane after marginalisation over $\h$ and $\bi$.
The contours denote the 68, 90,
95 and 99 per cent confidence regions.}
\label{results2}
\end{figure*}

\begin{figure*}
\centering
{\epsfig{file=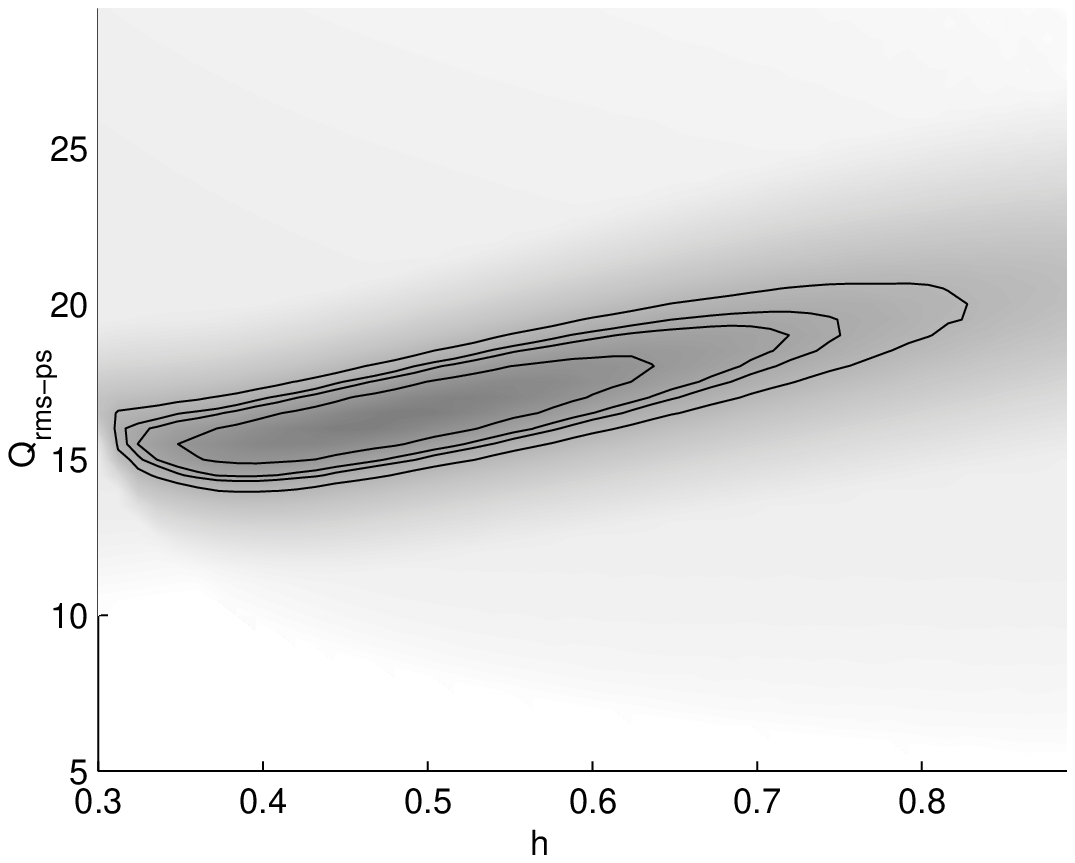,width=12cm}}
\caption
{The joint CMB+IRAS likelihood function in the
\{$\h$, $Q$\}-plane after marginalisation over $\omegam$ and $\bi$.
The contours denote the 68, 90, 95 and 99 per cent confidence regions.}
\label{results3}
\end{figure*}

The complementary nature of the two data sets is demonstrated in
Fig.~\ref{results1}, which shows likelihood contours in
the \{$\omegam$, $\h$\}-plane after marginalising over $Q$ and $\bi$.
The fundamental CMB-side degeneracy in 
\{$\omegam$, $\h$\} is seen in the flat trough running across 
Fig.~\ref{results1}a. The IRAS degeneracy is in a different direction 
(Fig.~\ref{results1}b) and the two data sets agree well in the region
where the lines of degeneracy overlap. Combining the two data sets
breaks the degeneracy and leads to a well defined joint optimum 
(Fig.~\ref{results1}c).
The joint CMB plus IRAS likelihood function in the 
\{$\omegam$, $Q$\}-plane and \{$\h$, $Q$\}-plane is plotted in
Figs~\ref{results2} \& \ref{results3} respectively.
In each case, a marginalisation has been performed over the other two parameters.

The joint likelihood (equation~\ref{jointlike}) was maximized with
respect to the 4 free parameters (equation~\ref{alphajoint}) using
standard optimisation techniques (\cite{press92}) and the best fit
parameters are shown in Table~\ref{valuetable}.
\begin{table}
\centering
\begin{tabular}{@{}lc r@{.}l @{$\,<\,$} c @{$\,<\,$} r@{.}l}
             &Free parameters           \\
$\omegam$    &$0.39$    &  $0$ & $29$ &$\omegam$&$0$&$53$\\
$h$          &$0.53$    &  $0$ & $39$ &$h$      &$0$&$58$ \\
$Q$ ($\mu$K) &$16.95$   &  $15$& $34$ &$Q$      &$17$&$60$\\
$\bi$        &$1.21$    &  $0$ & $98$ &$\bi$    &$1$&$56$\\
\\
             &Derived  parameters \\
$\Omega_b$   &$0.085$\\
$\s8$        &$0.67$\\
$\sigma_{8,{\rm iras}}$        &$0.81$\\
$\Gamma$     &$0.15$\\
$\betai$     &$0.47$\\
Age (Gyr)    &$16.5$\\
\\
\end{tabular}
\caption{Parameter values at the joint optimum.
For the free parameters the  
68\% confidence limits are shown,
calculated for each parameter by marginalising the likelihood over the other 
variables.}
\label{valuetable}
\end{table}
For this set of parameters, we find the values of the reduced $\chi^2$
for the IRAS and CMB data respectively to be 1.18 and 1.03, confirming
that both data-sets agree well with the models used.  Taking the CMB
and IRAS data together the total reduced $\chi^2$ is 1.16. Recalculating
the joint optimum using the simpler formula $\Gamma = \omegam\,\h$ in
place of that in equation~\ref{eqngamma} had little effect. Further,
the optimum was robust to changes in IRAS $l_{\rm max}$ in the range
$4 \le l_{\rm max} \le 10$.

\begin{figure*}
\centering
{\epsfig{file=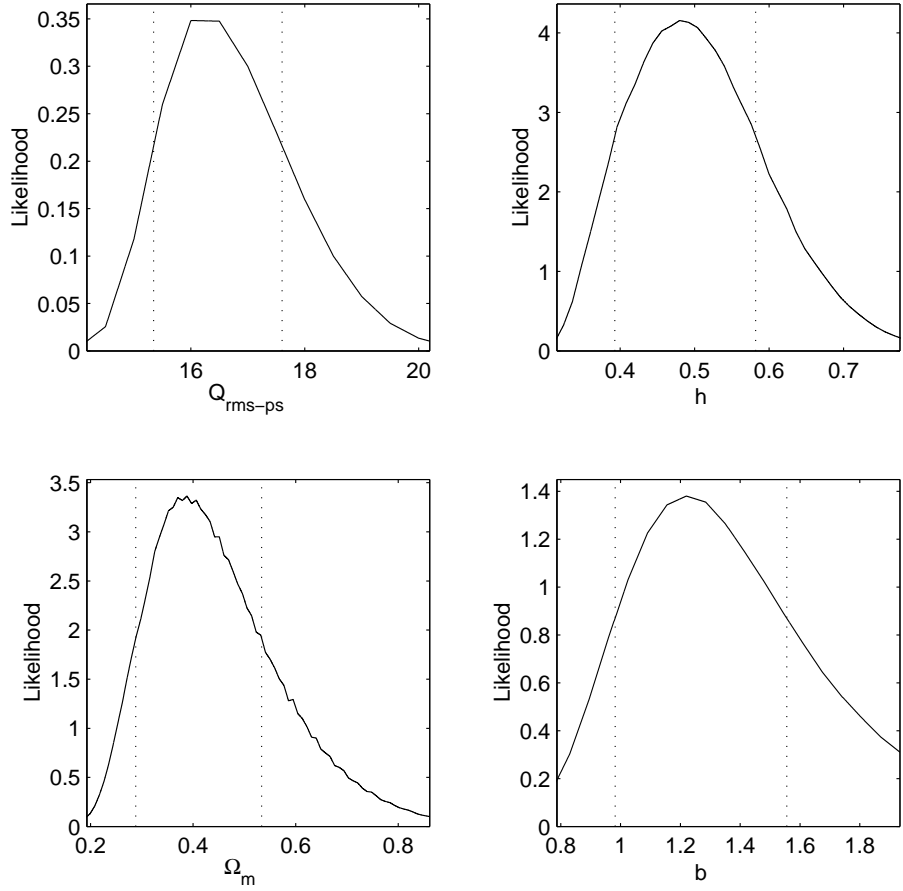,width=12cm}}
\caption
{The one-dimensional marginalised probability distributions for
each of the four parameters. The vertical dashed lines denote the 68\%
confidence limits. The horizontal plot limits are at the 99\%
confidence limits.} 
\label{margplots}
\end{figure*}

To obtain 68 per cent confidence limits on each of the free parameters
it is necessary to marginalise over the remaining free parameters. To
achieve this we 
evaluated the joint likelihood function on a 4-dimensional grid of
parameter values. The range of values and number of grid points
in each direction were $5<Q<30$, $50$ steps; $0.3<\h<0.9$, $50$ steps; 
$0.1<\omegam<1.0$, $100$ steps; $0.7<\bi<2.0$, $20$ steps.
For each parameter, the corresponding one-dimensional marginalised
probability distribution was
calculated by integrating the likelihood function over the other
variables. The marginalised distribution for each parameter is
shown in Fig.~\ref{margplots}, in which the dashed vertical lines denote
the 68\% confidence limits quoted in Table~\ref{valuetable}.
In general, the peak of the one-dimensional
probability distribution for each variable will differ from the
global optimum across all parameters. However, for all four variables
in this system, the two values are found to be extremely close.

In addition 
we evaluated the covariance matrix at the joint optimum. The
covariance matrix is simply the inverse of the Hessian
at the joint optimum. The Hessian is given by 
$ { {\partial^2{\rm ln}\left({\cal L}_{\rm joint}\right)} /
  ({\partial \alpha_i \partial \alpha_j}) } $
for pairs of parameters $\alpha_i$ and $\alpha_j$ 
and is evaluated using a standard
central-difference algorithm. Taking the square-root of the diagonal
elements of the covariance matrix, the standard errors on the
parameters are found to be $\Delta Q = 0.67$ $\mu$K, $\Delta\h =
0.03$, $\Delta\omegam=0.04$ and $\Delta\bi=0.1$. Comparing these with
the marginalised errors quoted in Table~\ref{valuetable} we note that
the marginalised errors are consistently larger, as expected.
By rescaling the covariance matrix so that its diagonal elements equal
unity, we obtain the correlation matrix shown in Table~\ref{covartable}.
We see that the most strongly correlated parameters are 
$\omegam$ and $h$.
\begin{table}
\centering
\begin{tabular}{@{}lrrrr}
          &$Q$       &$\h$      &$\omegam$    &$\bi$  \\
$Q$       &$1.00$    &$0.40$    &$-0.46$      &$-0.49$\\
$\h$      &$0.40$    &$1.00$    &$-0.82$      &$-0.57$\\
$\omegam$ &$-0.46$   &$-0.82$   &$1.00$       &$0.09$ \\ 
$\bi$     &$-0.49$   &$-0.57$   &$0.09$       &$1.00$ \\
\\
\end{tabular}
\caption{Parameter correlation matrix at the joint optimum}
\label{covartable}
\end{table}
%


\section{Discussion}

The results of this joint optimisation are in reasonable agreement
with other current estimates. 
The relatively low value of $\omegam \approx 0.4$ is close to that found by
others (\cite{white93,bahcall97}), and is in line with
recent supernovae results (\cite{perlmutter98}). However, given 
the assumption of a
flat universe, it requires a very high cosmological constant 
($\omegal = 0.6$). Gravitational lensing measurements 
have constrained $\omegal < 0.7$ (\cite{kochanek96}, \cite{falco}).
Our value for the Hubble constant, 
$\h = 0.53$, agrees well with several other measurements
(\cite{sugiyama95,lineweaver97}), but
falls at the low end of the generally accepted range from local
measurements (\cite{freedman94}). Assuming the nucleosynthesis
constraint $\Omega_b h^2 = 0.024$ (\cite{tfb96}, \cite{shf98}),
the optimal baryon density is found to be $\omegab = 0.085$.
Our value for the combination $\s8 \omegam^{0.6} = 0.38$ 
is lower than the one derived from 
measurements from the peculiar velocity field, 
$\sigma_8 \omegam^{0.6} \approx 0.8$ (\cite{zehavi98}).
Our values are closer to the combination
derived from cluster abundance  
$\sigma_8 \omegam^{0.5} \approx 0.5 $ (\cite{eke96}, \cite{bahcall97}).
Finally, for spatially-flat universes the time since
the Big Bang 
for the values of our $\omegam$ and $h$ 
at the joint optimum is
$16.5$ Gyr. 

On the IRAS side, $\betai = 0.47$ is in agreement with several other
measurements (\cite{strauss89,schlegel95,willick97}), although there are
other measurements  which place $\betai$ much higher
(\cite{dekel93,sigad97}).  \cite{willick97} discuss
the discrepancies between the various measurement techniques, and why
they lead to such distinct results. Finally, the IRAS mass-to-light
bias is seen to be slightly greater than unity ($\bi = 1.21$), suggesting
that the IRAS galaxies (mainly spirals) are reasonable (but not perfect) 
tracers of the underlying mass distribution.
We also note that our joint IRAS+CMB 
optimal values for $\sigma_{8,{\rm iras}}, \beta_{{\rm iras}}$
and $\Gamma$ (Table 1) are in perfect agreement with 
the values derived from IRAS alone 
(FSL, \cite{fisher_paris}).
However at fixed $\sigma_{8, {\rm iras}} = 0.69$ 
based on the IRAS correlation function FSL found 
a higher $\beta_{{\rm iras} } = 0.94 \pm 0.17 $
and $\Gamma = 0.17 \pm 0.05 (1-\sigma)$. 
We emphasise that the naive linear biasing 
should be generalised to more realistic scenarios (e.g. \cite{dl98}).

As discussed in Section~\ref{results}, the total CMB+IRAS reduced
$\chi^2$ at the joint optimum is 1.16, indicating that the optimum
model is a good fit to both data-sets. We may compare this value with
that obtained by \cite{gs98}.
Using several  data-sets \cite{gs98} found
the reduced $\chi^2$ for $\Lambda$CDM  to be 1.9, 
as opposed to 
a value of 1.2 for their `best' model of Cold+Hot Dark matter (CHDM).
However, using only the 
CMB and IRAS $\chi^2$ values for $\Lambda$CDM in
their Table 3, the total reduced $\chi^2$ is found to be 1.00 as
compared to a value of 0.95 for CHDM. Thus, using only the CMB and
IRAS data-sets, $\Lambda$CDM is found to fit the observations as
well as CHDM, and the results quoted by \cite{gs98}
are consistent with those presented here.

The near future will see a dramatic increase in LSS data
(e.g.\ the PSCZ, SDSS, 2dF surveys) and detailed measurements of 
the CMB fluctuations on sub-degree scales (e.g.\ from the Planck
Surveyor and MAP satellites). 
These will allow more accurate parameter estimation and the untying 
of the various parameters held fixed in the present work.


\subsection*{Acknowledgments}
We would like to thank E. Gawiser, J. Felten, J. Silk, G. Steigman,
K. Wu and I. Zehavi for helpful discussions. Matthew Webster and Sarah
Bridle acknowledge PPARC studentships.  Gra\c ca Rocha acknowledges a
NSF grant EPS-9550487 with matching support from the State of Kansas
and from a K*STAR First award.



\label{lastpage}


\begin{thebibliography}{}
\bibitem[Bagla 1998]{bagla98}
        Bagla, J., S., 1998, MNRAS, in press, astro-ph/9711081) 
\bibitem[Bahcall, Fan and Cen 1997]{bahcall97} 
	Bahcall N.A., Fan, X., \& Cen, R. 1997, ApJ, 485, L53
\bibitem[Baker \etal 1998]{cat2} 
	Baker J.C. \etal, 1998, MNRAS, submitted
\bibitem[Bardeen \etal 1986]{bbks86} 
	Bardeen J.M., Bond J.R., Kaiser N. \& Szalay A.S., 1986,
	ApJ, 304, 15
\bibitem[Bennett \etal 1996]{bennett}
        Bennett C.L. \etal, 1996, ApJ., 464, L1
\bibitem[Blanton \etal 1998]{blanton98}
         Blanton, M., Cen, R., Ostriker, J.P., \& Strauss, M.A., 
         1998, astro-ph/9807029
\bibitem[Bond 1995{\it a}]{bond95a} 
	Bond J.R., 1995, {\it Cosmology and Large Scale Structure}
	ed. Schaeffer, R.
	Elsevier Science Publishers, Netherlands, Proc. 
        Les Houches School, Session LX,
	August 1993
\bibitem[Bond 1995{\it b}]{bond95b}
	Bond J.R., 1995, Astrophys. Lett. and Comm., 32, 63
\bibitem[Bond \etal 1995{\it c}]{bond95c} Bond J.R., Davis
	R.L. \& Steinhardt P.J., 1995,  Astrophys. Lett. and Comm., 32, 53
\bibitem[Carroll, Press \& Turner 1992]{cpt92} 
	Carroll S.M., Press W.H. \& Turner E.L., 1992, 
        Ann. Rev. Astron. \& Astrophy., 30, 499
\bibitem[Cheng \etal 1994]{msam1} 
	Cheng E.S. \etal, 1994, ApJ., 422, L37
\bibitem[Cheng \etal 1996]{msam2} 
	Cheng E.S. \etal, 1996, ApJ., 456, L71
\bibitem[Cole \etal 1995]{cole95} Cole S., 
	Fisher K.B. \& Weinberg D.H., 1995,  MNRAS, 275, 515
\bibitem[De Bernardis \etal 1994]{argo} 
	De Bernardis P. \etal, 1994, ApJ., 422, L33
\bibitem[Dekel \etal 1993]{dekel93} 
	Dekel A., Bertschinger E., Yahil A., Strauss M., Davis M. \&
        Huchra J., 1993, ApJ, 412, 1
\bibitem[Dekel and Lahav 1998]{dl98}
        Dekel, A. \& Lahav, O., submitted to ApJ, astro-ph/9806193 
\bibitem[Efstathiou, Bond \& White 1995]{ebw92} 
	Efstathiou G., Bond J.R. \& White S.D.M., 1992, MNRAS, 258, P1
\bibitem[Eisenstein, Hu \& Tegmark 1998]{eht98}
        Eisenstein, D.J., Hu, W. \& Tegmark, M., 1998, astro-ph/9807130
\bibitem[Eke, Cole \& Frenk 1996]{eke96} 
        Eke, V.R., Cole, S., and Frenk, C.S., 1996, MNRAS, 282, 263
\bibitem[Falco, Kochanek \& Munoz 1998]{falco}
	Falco, E.E, Kochanek, C.S. \& Munoz, J.A., 1998, ApJ, 494, 47
\bibitem[Fisher 1994]{fisher_paris}
        Fisher, K.B., in proceedings of `Cosmic Velocity Fields', 
        Paris, eds. Bouchet et al., pg. 177, Editions Frontieres
\bibitem[Fisher, Scharf \& Lahav 1994]{fsl94}
        Fisher, K.B., Scharf, C.A. \& Lahav, O., 1994, MNRAS, 266, 219
\bibitem[Fisher \etal 1995]{fisher95}
        Fisher, K.B., Huchra, J.P., Strauss, M.A., Davis, M., Yahil A., 
        Schlegel D., 1995, ApJ, 100, 69
\bibitem[Fisher \& Nusser 1996]{fn96}
        Fisher K.B. \& Nusser A., 1996, MNRAS, 279, L1
\bibitem[Freedman \etal 1994]{freedman94}
        Freedman W., \etal,
        1994, Nature, 371, 757
\bibitem[Freudling et al. 1998]{zehavi98}
         Freudling et al., 1998, submitted to ApJ
\bibitem[Gawiser \& Silk 1998]{gs98}
        Gawiser E., \&  Silk J., 1998, Science, 280, 1405
\bibitem[Gundersen \etal 1995]{spole} 
	Gundersen J.O \etal, 1995, ApJ., 443, L57
\bibitem[Gutierrez 1997]{guti97} 
	Gutierrez C.M., 1997, ApJ., 483, 51
\bibitem[Hancock \etal 1994]{nature94} 
	Hancock S., \etal, 1994, Nature, 367, 333
\bibitem[Hancock \etal 1997]{me96}
	Hancock S., Gutierrez C.M., Davies R.D., Lasenby A.N., 
        Rocha G., Rebolo R., Watson R.A., Tegmark M., 1997, MNRAS,
        298, 505
\bibitem[Hancock \etal 1998]{sh97}
	Hancock S., Rocha G., Lasenby A.N., Gutierrez C.M., 1998,
        MNRAS, 294, L1
\bibitem[Heavens \& Taylor 1995]{ht95}
        Heavens A.F. \& Taylor A.N., 1995, MNRAS, 275, 483
\bibitem[Kaiser 1987]{kaiser87}
        Kaiser N., 1987, MNRAS, 227, 1
\bibitem[Kaiser \etal 1991]{kaiser91}
        Kaiser N., Efstathiou G., Ellis R., Frenk C., Lawrence A., 
        Rowan-Robinson M., Saunders W., 1991, MNRAS, 252, 1
\bibitem[Kochanek 1996]{kochanek96} 
	Kochanek C.S., 1996, ApJ, 466, 638
\bibitem[Lahav \etal 1994]{lahav94}
        Lahav, O., Fisher, K.B., Hoffman, Y., Scharf, C.A, \& Zaroubi, S., 
        1994, ApJL, 423, L93
\bibitem[Lahav 1996]{lahav96} 
	Lahav O., 1996, Helv. Phys. Acta, 69, 388
\bibitem[Leitch \etal 1998]{ovro}
        Leitch E.M., Readhead A.C.S., Pearson T.J., Myers S.T., \& Gulkis
        S., 1998, astro-ph/9807312
\bibitem[Lineweaver \etal 1997]{lineweaver97} 
	Lineweaver C.H., Barbosa D., Blanchard A. \& Bartlett J.G.,
	1997, Astron. \& Astrophy., 322, 365
\bibitem[Netterfield \etal 1997]{sask} 
	Netterfield C.B., Devlin M.J., Jarosik N., Page L., 
        Wollack E.J., 1997, ApJ, 474, 47
\bibitem[Peebles 1980]{peebles80}
        Peebles, P.J.E., 1980, {\it The Large-Scale Structure of the Universe},
        (Princeton:  Princeton University Press)
\bibitem[Pen 1998]{pen98}
        Pen, U.-L., 1998, astro-ph/9711180
\bibitem[Perlmutter \etal 1998]{perlmutter98}
        Perlmutter S., \etal, 
        1998, Nature, 391, 51
\bibitem[Platt \etal 1997]{platt}
        Platt S.R., Kovac J., Dragovan M., Peterson J.B., \& Ruhl J.E., 
	1997, ApJ, 475, L1
\bibitem[Press \etal 1992]{press92} 
	Press, W.H., Teukolsky, S.A., Vetterling, W.T., 
        \& Flannery, B.P., 1992, {\it Numerical Recipes (Second Edition)}
        (Cambridge: Cambridge University Press)
\bibitem[Rocha \etal 1998]{gr}
	Rocha, G., Hancock, S., Lasenby, A.N., \& Gutierrez, C.M. 
        in preparation.
\bibitem[Ruhl \etal 1995]{python} 
	Ruhl J.E., Dragovan M., Platt S.R., Kovac J., \& Novak G.,
	1995, ApJ., 453, L1
\bibitem[Scaramela \& Vittorio 1990]{cosvariance2} 
	Scaramela N., \& Vittorio N., 1990, ApJ., 353, 372
\bibitem[Scaramela \& Vittorio 1993]{cosvariance1} 
	Scaramela N., \& Vittorio N., 1993, ApJ., 411, 1
\bibitem[Schlegel 1995]{schlegel95} 
	Schlegel D., 1995, Ph. D. Thesis, University of California, Berkeley
\bibitem[Scott, Srednicki and White 1994]{samvariance} 
	Scott D., Srednicki M., \&  White M., 1994, ApJ., 241, L5
\bibitem[Scott \etal 1996]{cat} 
	Scott, P.F. \etal, 1996, ApJ., 461, L1
\bibitem[Seljak \& Zaldarriaga 1996]{seljak} 
	Seljak U., \& Zaldarriaga M., 1996, ApJ, 469, 437
\bibitem[Sigad \etal 1998]{sigad97} 
	Sigad Y., Dekel A., Strauss M.A. \& Yahil A., 1998, 
        495, 516
\bibitem[Smoot \etal 1992]{smoot} 
	Smoot, G.F. \etal, 1992, ApJ., 396, L1
\bibitem[Steigman, Hata \& Felten 1998]{shf98}
         Steigman, G., Hata, N. \& Felten, J.E., 1998, 
         astro-ph/9708016 
\bibitem[Strauss 1989]{strauss89} 
	Strauss M.A., Ph. D. Thesis, University of California, Berkeley
\bibitem[Strauss \& Willick 1995]{sw95} 
	Strauss, M.A., \& Willick, J.A., 1995, Phys Rev, 261, 271
\bibitem[Sugiyama 1995]{sugiyama95} 
	Sugiyama N., 1995, ApJ Supp., 100, 281
\bibitem[Tanaka \etal 1996]{max} 
	Tanaka, S.T. \etal, ApJ, 468, L81
\bibitem[Tegmark \& Peebles 1998]{tp98}
        Tegmark, M., \& Peebles, P.J.E. 1998, astro-ph/9804067
\bibitem[Tytler, Fan \& Burles 1996]{tfb96}
        Tytler D., Fan X.M. \& Burles S., 1996, Nature, 381, 207
\bibitem[White \etal 1993]{white93}
        White \etal, 1993, Nature, 366, 429
\bibitem[Willick \etal 1997]{willick97}
        Willick J.A., Strauss M.A., Dekel A. \& Kolatt T., 1997, 
	ApJ, 486, 629
\bibitem[Yahil \etal 1991]{yahil91}
        Yahil, A., Strauss, M.A., Davis, M., \& Huchra, J.P., 1991, ApJ, 
        372, 380
\end{thebibliography}
\end{document}